\documentclass[12pt]{article}

\usepackage{url}
\usepackage{latexsym}
\usepackage{amsmath}
\usepackage{booktabs}
\usepackage{amssymb}
\usepackage{relsize}
\usepackage{graphicx}
\usepackage{epsfig}
\usepackage{ccaption}

\usepackage{a4wide}

\newcommand{\startappendix}{
\setcounter{section}{0}
\renewcommand{\thesection}{\Alph{section}}}

\newcommand{\Appendix}[1]{
\refstepcounter{section}
\vspace{10mm}
\pagebreak[3]
\setcounter{equation}{0}
\begin{flushleft}
{\large\bf Appendix \thesection: #1}
\end{flushleft}}

\newcommand{\be}{\begin{equation}}
\newcommand{\ee}{\end{equation}}
\newcommand{\bea}{\begin{eqnarray}}
\newcommand{\eea}{\end{eqnarray}}
\newcommand{\nn}{\nonumber\\}

\def\d{d}
\def\ln{\mathrm{ln}}
\def\ds{\displaystyle}

\def\a{\alpha}

\def\Om{\Omega}

\def\p{\phi}
\def\D{\Delta}

\def\L{\mathcal{L}}
\def\r{\rho^2_0}

\def\thus{\Rightarrow}

\title{\bf Extracting spacetimes using the AdS/CFT conjecture}

\author{Samuel Bilson\footnote{s.c.bilson@durham.ac.uk}\\ \\
{\it Department of Mathematical Sciences,}
\\
{\it Science Laboratories, South Road, Durham DH1 3LE, United Kingdom}}

\title{\bf Extracting spacetimes using the AdS/CFT conjecture}

\begin{document}
\begin{titlepage}
\maketitle

\maketitle

%Abstract
\begin{abstract}
We present analytic methods for extracting a class of bulk geometries given information of certain physical quantities in the boundary CFT. More specifically we look at singular correlators and entanglement entropy in the CFT to provide information of null and spacelike geodesics repectively in the bulk. We show that static spherically symmetric, asymptotically AdS spacetimes which do not admit null circular orbits can be fully recovered, and that any spacetime can be recovered up to the local maximum of the potential. We provide analytical and numerical examples to verify the methods used.     
\end{abstract}
\thispagestyle{empty}
\end{titlepage}

\section{Introduction}

The AdS/CFT correspondence relates two objects: 1) \textit{Type IIB superstring theory on} $\mathrm{AdS}_5\times \mathrm{S}^5$ and 2) $\mathcal{N}=4$ \textit{SU(N) super-Yang-Mills theory in} $3+1$ \textit{dimensions} \cite{MAGOO, Witten1, Maldacena}. The AdS/CFT correspondence is holographic, since it relates theories living in different dimensions. One consequence is that a $d+1$ dimensional asymptotically AdS gravitational theory is ``dual'' to a $d$ dimensional field theory which lives on the boundary of this spacetime. For example, it is well known that a spacetime containing a large black hole has a thermal CFT dual, where its temperature is the Hawking temperature \cite{Witten}. 

\paragraph{}To gain a deeper understanding of many issues that arise in quantum gravity, it has been useful to probe the AdS/CFT correspondence. Many problems of qauntum gravity manifest themselves in what happens beyond the horizon of black holes. Much research has been done into extracting information from inside the horizon of black holes from knowledge of the boundary field theory correlators \cite{Bal, Louko, Maldacena1, Kraus, Levi, Fid, Fest}. This has motivated original work by V. Hubeny et.al.\cite{hubeny}, to investigate singularities in the boundary correlation functions. They discovered that \textit{a boundary correlation function} $G(x,x^\prime) = \langle \mathcal{O}(x)\mathcal{O}(x^\prime) \rangle$ \textit{is singular if there exists null geodesics connecting $x$ and $x^\prime$}. The singularities arising from such a relation are known as ``bulk-cone singularities''. This relationship was then used to answer questions such as: \textit{can we see horizon formation directly in the gauge theory}?

\paragraph{}Complementary to investigating local properties of the CFT, one can also study non-local quantities. Then one can ask questions about what physical quantities would they correspond to in the bulk theory . One such quantity is the entanglement entropy $S_A$. In particular, it has beed  recently shown \cite{Ryu, Ryu1} that \textit{the entropy $S_A$ in a $d+1$ dimensional CFT can be determined from the d dimensional minimal surface $\gamma_A$} (in the bulk AdS) \textit{whose boundary is given by the $d-1$ dimensional manifold $\partial\gamma_A=\partial A$}. The entropy is given by applying the usual area/entropy relation. In the case of $\mathrm{AdS}_3$, the minimal surface is one dimensional, and is in fact a spacelike geodesic with endpoints on the boundary. The entropy in this case is just derived from the ``area'' of the spacelike geodesic, i.e. the proper length $\mathcal{L}$.  

\begin{figure}[htbp]
\label{AdS}
\unitlength1cm \hfil
\begin{picture}(16,16)
 \epsfxsize=6cm \put(5,8){\epsffile{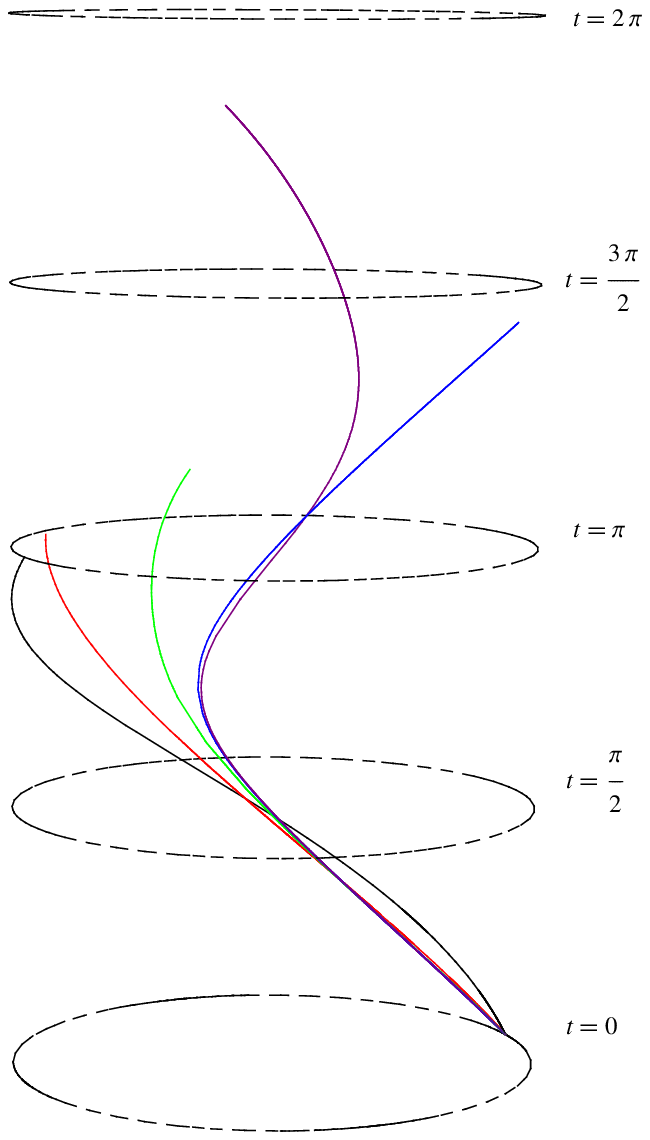}}
 \epsfxsize=9cm \put(0,1){\epsffile{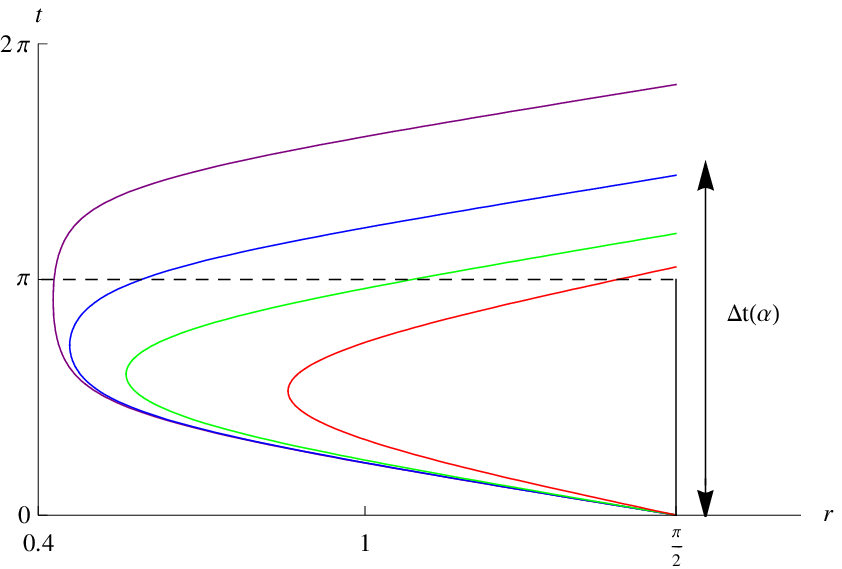}}
 \epsfxsize=8cm \put(9,0.25){\epsffile{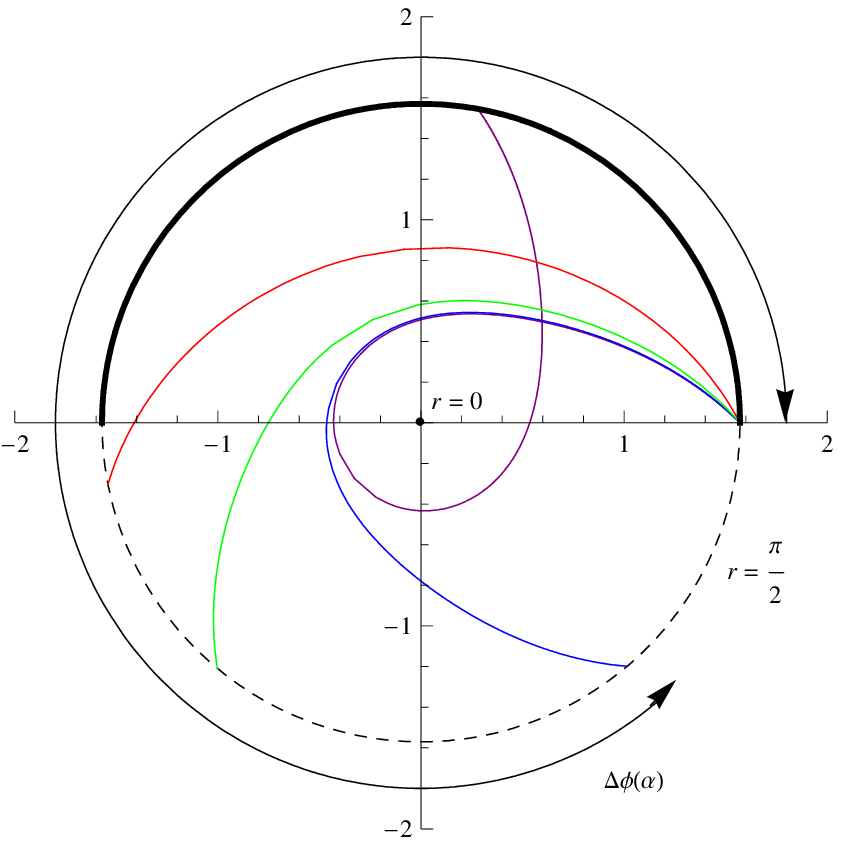}}

\put(7.75,7.5){(a)}
\put(3.75,0){(b)}
\put(12,0){(c)}
\end{picture}
\caption{$(a)$ is a $2+1$-dimensional plot of boundary to boundary null geodesics in an AdS-black hole geometry with metric $\d s^2=-f(r)\,\d t^2+\frac{\d r^2}{f(r)}+r^2\d\Om^2_n$ and $f(r)=1+r^2-\frac{1}{10r^2}$, where $n-1$ angular directions have been projected out. To plot the null trajectories, the spacetime has been compactified using the coordinate transformation $r=\mathrm{tan}(\tilde{r})$, thus the boundary is a timelike cylinder with radius $\tilde{r}=\frac{\pi}{2}$. $(b)$ illustrates the quantity $\D t(\a)$ via projection of the null geodesics on the $r$-$t$ plane, where $\a=\frac{E}{J}$ parameterises the null curves. $(c)$ illustrates the quantity $\D\p(\a)$ via projection of the null geodesics on the $r$-$\p$ plane. Notice that the boundary null curve ($\a=1$), we have $\D t(\a)=\D\p(\a)=\pi$, which is as expected from a pure AdS calculation. (Colour coding for values of $\a$: $\{1,1.3,1.7,1.85,1.87\}=\{$Black, Red, Green, Blue, Purple$\}$)}

\end{figure}

\paragraph{}Using the relation shown in \cite{hubeny}, numerical work was done to reconstruct certain static, spherically symmetric bulk metrics given the location of bulk-cone singularities \cite{hammer1, hammer2}. This paper complements the previous work by attempting to reconstruct the bulk metrics using inversion techniques and thus provide an \textit{exact solution}. This is done by noting that the location of the boundary singularities can be expressed in terms of two functions, $\D\p(\a)$ and $\D t(\a)$, where if $G(x,x^\prime)$ is some singular correlator between the points $x=(t,\Om)$ and $x^\prime=(t^\prime,\Om^\prime)$, then $t-t^\prime=\D t$ and $\Om-\Om^\prime\propto\D\p$ (see figure.\ref{AdS}). These two functions can in turn be completely determined by the bulk metric, thus an inversion is required to determine the bulk metric.

\paragraph{}In this paper we will consider two types of spacetime. In section 2 we will descibe static spherically symmetric spacetimes described by one function $f(r)$. We will then introduce the inversion technique following from the ideas set out in the introduction. Analytical and numerical examples are then given to test the validity of the inversion. In section 3 we generalise to static spherically symmetric spacetimes decribed by two functions $f(r)$ and $h(r)$. It will be noted that information of the endpoints of null geodesics will not be enough to fully determine the metric. However, supplementing the bulk-cone singularity data with the entanglement entropy for arbitrary CFT regions will turn out to suffice to extract the bulk metric. Since entanglement entropy in the two dimensional CFT can determine the proper length $\mathcal{L}$ of spacelike geodesics, and the proper length is expressed in terms of the metric, it will be shown that this is enough information in the specific case of an $\mathrm{AdS}_3$ bulk to completely determine the metric \textit{exactly}. Analytical and numerical examples are again given to test the validity of, in this case, two inversions. In section 4 we will give a discussion of the results. It will be noted that although we have extracted spherically symmetric asymptotically AdS spacetimes only in the case of 3 dimensions, it is completely valid for any dimensions. We will also discuss limitations of the inversion, and future areas that can be explored.

\newpage 

\section{Static spherically symmetric spacetimes decribed by one funcion $f(r)$}

An important class of static spherically symmetric spacetimes in $n+2$ dimensions have metrics, in global coordinates, of the form
\be
\label{eq:AdS}
\d s^2 = -f(r)\,\d t^2 + \dfrac{\d r^2}{f(r)} +r^2\,\d \Om^{2}_n
\ee
Putting the condition that as $r\rightarrow\infty$, $f(r)\rightarrow r^2+1$ restricts the spacetime to be asymptotically AdS. If we set $f(r)=1+r^2-p(r)$, then for the spacetime to be physically meaningful, $p(r)$ must be $O(r^n)$ where $n\leqslant -2$.

\paragraph{}Fixing one endpoint for all null geodesics on the boundary, calculation of the angular, $\D\p(\a)$, and time, $\D t(\a)$ (see figure. \ref{AdS}), separation of null geodesics (labelled by the ratio $\a=\frac{E}{J}$), with endpoints on the boundary (see Appendix A) yeilds the two expressions

\be
\D t(\a) = 2\int^{\infty}_{r_{min}(\a)}\frac{{\a}\,\d r}{f(r)\sqrt{{\a}^2 - \frac{f(r)}{r^2}}}
\ee
and
\be
\label{dphi}
\D\p(\a) = 2\int^{\infty}_{r_{min}(\a)}\frac{\d r}{r^2\sqrt{\a^2 - \frac{f(r)}{r^2}}}
\ee
where $r_{min}(\a)$ is the minimum radius reached in the bulk by a null geodesic, and is defined by $\dot{r}\mid_{r_{min}(\a)} = 0$ 

\paragraph{}

Examining the two terms, there is a relation between $\D\p(\a)$ and $\D t(\a)$ (see Appendix B for derivation) which goes as,

\be
\label{dadph}
\a\dfrac{\d}{\d\a}\D t(\a) = \dfrac{\d}{\d\a}\D\p(\a)
\ee

Although it appears that we have two functions worth of data to extract $f(r)$, they are not independent from the equation above. Thus we will use $\D\p(\a)$ only for the extraction.

\subsection{Extracting $f(r)$}

Examining the definition for $\D\p(\a)$, it is first noted that this is an integral equation of the first kind. The simplest equation of this form is $a(x)=\int^x_0b(r)\,\d r$, which has the trivial solution $b(r)=a^\prime(r)$. The extraction of $f(r)$ is complicated by the fact that equation \eqref{dphi} is non-linear in $f(r)$. Non-linearity in any type of equation makes finding an analytic solution much harder, so it would be useful to find a quantity which is linear in the integrand and invert that instead. This can be done by identifying linear equations which control the trajectory of null geodesics. One such equation is the ``energy equation''. From Appendix \ref{null} we have the equation for $\dot{r}$,
 
\be
\dot{r}^2 = \a^2 - \dfrac{f(r)}{r^2}
\ee

By inspection, we can identify the potential $V(r)=\dfrac{f(r)}{r^2}$, to get,

\be
\dot{r}^2+V(r)=\a^2
\ee

Instead of integrating over the quantity $r$, let us perform a substitution to integrate over the quantity $V(r)$ instead. If we let  $r_{min}(\a) = x$, we see an immediate simplification for the expression of $\D\p(\a)$,

\be
\D\p\left(\sqrt{V(x)}\right) = 2\int^{\infty}_x\dfrac{\d r}{r^2\sqrt{V(x) - V(r)}}
\ee

Now if we perform the substitution by letting $V(r)=V$, we have,
\be
\D\p\left(\sqrt{V(x)}\right) = \int^{V(x)}_{V(\infty)}\dfrac{p(V)\,\d V}{\sqrt{V(x) - V}},\quad\text{where}\quad p(V)=-2\,\dfrac{\d r}{\d V}\left[V^{-1}(V)\right]^{-2}
\ee
\paragraph{}We can now see that the integrand is linear in $p(V)$. To make the integral look nicer, we can let $V(x)=V_{max}$, and applying the condition of asymptotically AdS, we have,
\be
\label{pV}
\D\p\left(\sqrt{V_{max}}\right) = \int^{V_{max}}_{1}\dfrac{p(V)\,\d V}{\displaystyle\sqrt{V_{max} - V}}
\ee

Equation \eqref{pV} is in fact a Volterra equation of the First Kind. It is a particular form of the {\it generalized Abel equation} \cite{Abel}, which is of the form
\be
\int^x_a\dfrac{y(t)}{(x-t)^\lambda}\,\d t=f(x),\quad 0<\lambda<1.
\ee

and has solution,
\be
y(t)=\dfrac{\mathrm{sin}(\pi\lambda)}{\pi}\dfrac{\d}{\d t}\int_a^t\dfrac{f(x)\,\d x}{(t-x)^{1-\lambda}}
\ee

Applying this to equation \eqref{pV}, we have,
\be
p(V)=\dfrac{1}{\pi}\dfrac{\d}{\d V}\int^{V}_1\dfrac{\D\p\left(\ds\sqrt{V_{max}}\right)\,\d V_{max}}{\displaystyle\sqrt{V-V_{max}}}
\ee

Now substituting back in for the original variables,
\be
-2\,\dfrac{\d r}{\d V}\,r^{-2}=\dfrac{2}{\pi}\dfrac{\d}{\d V}\int^{\sqrt{V}}_1\dfrac{\a\,\D\p(\a)}{\displaystyle\sqrt{V-\a^2}}\,\d\a
\ee

\be
\label{exfr}
\Rightarrow \dfrac{1}{r}=\dfrac{1}{\pi}\int^{\sqrt{V(r)}}_1\dfrac{\a\,\D\p(\a)}{\displaystyle\sqrt{V(r)-\a^2}}\,\d\a
\ee

From equation \eqref{exfr}, in principle one can solve the integral for any $\D\p(\a)$ and then rearrange to find $V(r)$ and thus $f(r)$.

\subsection{Analytic extraction using equation \eqref{exfr}: Pure AdS case}

Now that we have an integral inversion technique for extracting $f(r)$, we can test it analytically. We will choose the simplest case of pure AdS in global coordinates, i.e. $f(r)=1+r^2$.

\paragraph{}From equation \eqref{dphi} we get $\D\p(\a)=\pi$ in pure AdS, thus using equation \eqref{exfr} we have,

\bea
\dfrac{1}{r}&=&\int^{\sqrt{V(r)}}_1\dfrac{\a}{\displaystyle\sqrt{V(r)-\a^2}}\,\d\a\nn
\Rightarrow \dfrac{1}{r}&=&\sqrt{V(r)-1}\nn
\Rightarrow f(r)&=&r^2V(r)=1+r^2\quad\text{as expected}\nn
\eea

\subsection{Numerical extraction using equation \eqref{exfr}}

Of course not all integrals can be evaluated analytically, but we can solve them numerically, which means we can test the extraction of a more complicated $f(r)$. Firstly we can consider a spacetime for which the position of the local maximum of $V(r)$ is at $r\leqslant 0$:

\begin{figure}[htbp]
\unitlength1cm \hfil
\begin{picture}(5,5)
 \epsfxsize=7cm \put(-1,0){\epsffile{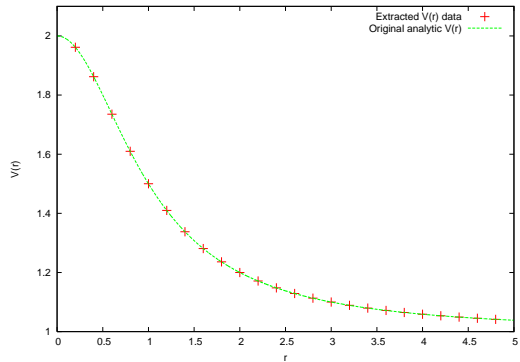}}
\end{picture}
\caption{Extraction of $V(r)$ where $f(r)=1+r^2-\dfrac{1}{r^2+1}$. In this case we can extract the entire spacetime.}
\end{figure}

\newpage

Now we can try a more complicated $f(r)$ by adding some Gaussian bumps:

\begin{figure}[htbp]
\unitlength1cm \hfil
\begin{picture}(7,7)
 \epsfxsize=9cm \put(-1,0){\epsffile{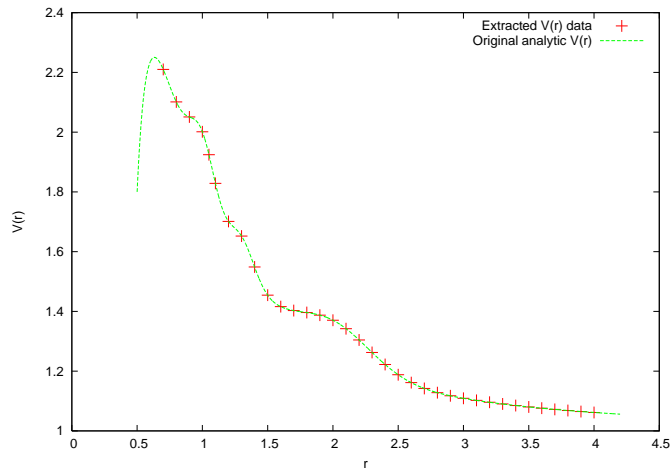}}
\end{picture}
\caption{Extraction of $V(r)$ where $f(r)=1+r^2-\dfrac{1}{5r^2}+r^2\displaystyle\sum_{i=1}^3\dfrac{\mathrm{A_i}}{\sqrt{2\pi}\sigma_i}\,e^{-\frac{(r-\mu_i)^2}{2{\sigma_i}^2}}$ and $\mathrm{A_i}=(0.05,0.03,0.1),\,\mu_i=(1,1.3,2)\,\,\text{with}\,\,\sigma_i=(0.1,0.1,0.3)$.}
\end{figure}

In this case we can only reproduce the spacetime down to a mimimum value of $r$ when $\frac{\d V(r)}{\d r}=0$. This is because null trajectories become circular orbits at this point. Null geodesics which go past this critical value will end up at the singularity and will thus never reach the boundary.

\paragraph{}In summary, we have found an analytic inversion method for extracting static spherically symmetric bulk spacetimes with one function given information of the bulk-cone singularities. Such spacetimes that can be extracted from this method include pure AdS and eternal black holes in AdS. The method was found by identifying linearity in the equations of motion for null gedesics, and use of Abel's equation. It was shown from graphical examples that any spacetime of this type can be extracted down to the local maximum of the potential. 

\paragraph{}In the next section we will expand on this method to static spherically symmetric spacetimes with two functions. The extraction method will be similar, but we will find that information of the bulk-cone singularities will not be enough to determine the metric down to the local maximum. 

\section{General static spherically symmetric spacetimes}

Metrics of the most general static, spherically symmetric spacetimes in $n+2$ dimensions are given by,

\begin{equation}
\label{frhr}
{d}s^2 = -f(r){d}t^2 + \dfrac{{d}r^2}{h(r)} +r^2{d}\Omega^2_n
\end{equation}

Applying the condition that the spacetime is asymptotically AdS, we let $f(r), h(r)\rightarrow 1+r^2$ as $r\rightarrow\infty$.

\paragraph{}Following the proceedure in Appendix \ref{null}, the equations of motion along null trajectories are,

\begin{equation}
\label{nullstar}
\dot{t} = \dfrac{\alpha}{f(r)},\quad\dot{\phi} = \dfrac{1}{r^2},\quad\dot{r}^2 = \dfrac{h(r)}{f(r)}\left[{\alpha}^2 - \dfrac{f(r)}{r^2}\right]
\end{equation}

which leads to the two modified quantities

\be
\D t(\a) = 2\int^{\infty}_{r_{min}(\a)}\frac{{\a}\,\d r}{\sqrt{f(r)\,h(r)\,({\a}^2 - \frac{f(r)}{r^2})}}
\ee
and
\be
\label{dphi1}
\D\p(\a) = 2\int^{\infty}_{r_{min}(\a)}\frac{\sqrt{f(r)}\,\d r}{r^2\sqrt{h(r)\,(\a^2 - \frac{f(r)}{r^2})}}
\ee

where the same definitions are used as in Section 2.

\subsection{Extracting $h(r)$}

In parallel with the extraction of section 2, we will consider the $\D\p(\a)$ function only, only now there are two functions to extract to fully determine the metric. We will start by considering the extraction of $h(r)$.

\paragraph{}It was noted in the introduction that a relationship has been found between the proper length $\L$ of boundary-to-boundary spacelike geodesics and the entangement entropy $S_A$ of the dual CFT. It will now be shown that given knowledge of $S_A$ and thus $\L$, we can extract $h(r)$.

\paragraph{}The expression of the proper length $\L$ for boundary-to-boundary static spacelike geodesics with the metric given in \eqref{frhr} (see Appendix \ref{spgeo}) is given by,

\begin{equation}
\L(J) = 2\int^{\rho_0}_J\dfrac{2r\,\d r}{\sqrt{h(r)\,(r^2 - J^2)}}
\end{equation}

If we massage the integral by letting $y(r) = -\dfrac{2r}{\sqrt{h(r)}}$, we get:
\begin{equation}
\L(J) = \displaystyle\int^J_{\rho_0}\dfrac{y(r)}{\sqrt{r^2 - J^2}}\,\d r
\end{equation} 

Notice this is a linear integral equation and is in fact a modified version of Abel's equation, which has solution,

\begin{equation}
y(r) = \dfrac{2}{\pi}\dfrac{\d}{\d r}\int^r_{\rho_0}\dfrac{J\L(J)}{\sqrt{J^2 - r^2}}\,\d J
\end{equation}
Substituting back in for $h(r)$, we have the final equation with which to extract $h(r)$ given $\L(J)$
\begin{equation}
\label{exhr}
\dfrac{1}{\sqrt{h(r)}} = \dfrac{1}{r\pi}\dfrac{\d}{\d r}\int^r_{\rho_0}\dfrac{J\L(J)}{\sqrt{J^2 - r^2}}\,\d J
\end{equation}

\subsubsection{Analytic extraction using equation \eqref{exhr}}

Now we can check the extraction with some trial functions. Since $h(r)$ behaves like $1+r^2$ for $r\rightarrow\infty$, we begin with the simplest case\footnote{Note there is a simpler case with which one can check the inversion, where $h(r)=1$, which is Euclidian space. In this case $\L(J)$ is just the length of a chord on a circle of radius $\rho_0$, i.e. $\L(J)=2\sqrt{\r-J^2}$. Then it is just an exercise in integral solving to show that this does indeed imply $h(r)=1$.} of $h(r)=1+r^2$ .

\paragraph{}We start by determining the proper length $\L$.

\bea
\L(J)&=&-2\int^J_{\rho_0}\dfrac{r\,\d r}{\displaystyle\sqrt{(1+r^2)(r^2-J^2)}}\\
x=r^2-J^2\thus\L(J)&=&\int^{\rho_0^2-J^2}_0\dfrac{\d x}{\ds\sqrt{x(x+1+J^2)}}\nn
\thus\L(J)&=&2\,\ln\left(\dfrac{\sqrt{\r+1}+\sqrt{\r-J^2}}{\ds\sqrt{1+J^2}}\right)\nn
\eea

Now we can determine $h(r)$, using equation \eqref{exhr} we have,

\be
\dfrac{1}{\sqrt{h(r)}} = \dfrac{1}{r\pi}\dfrac{\d}{\d r}\int^r_{\rho_0}\dfrac{J\L(J)}{\sqrt{J^2 - r^2}}\,\d J=\dfrac{1}{r\pi}\dfrac{\d}{\d r}I(r)
\ee
where
\be
I(r)=\int_{\rho_0}^r\dfrac{2J\,\d J}{\ds\sqrt{J^2-r^2}}\,\ln\left(\dfrac{\sqrt{\r+1}+\sqrt{\r-J^2}}{\ds\sqrt{1+J^2}}\right)
\ee 
Integrating by parts \cite{integrals}, gives
\be
I(r)=\int_{\r}^{r^2}\dfrac{\sqrt{x-r^2}}{1+x}\,\d x
+\int^{r^2}_{\r}\dfrac{\sqrt{x-r^2}}{\sqrt{\r-x}\,(\sqrt{\r-x}+\sqrt{\r+1})}\,\d x=I_1(r)+I_2(r)
\ee
where
\be
I_1(r)=-2\sqrt{\r-r^2}+2\sqrt{r^2+1}\,\mathrm{arctan}\left(\sqrt{\dfrac{\r-r^2}{1+r^2}}\right)
\ee
and
\be
I_2(r)=I_{21}(r)+I_{22}(r)+I_{23}(r)
\ee
where
\be
I_{21}(r)=2\sqrt{\r-r^2},\, I_{22}(r)=-\pi\sqrt{\r+1}\,\text{and}\,I_{23}(r)=\pi\sqrt{r^2+1}-2\sqrt{r^2+1}\,\mathrm{arcsin}\left(\sqrt{\dfrac{\r-r^2}{\r+1}}\right)
\ee
\bea
\thus I(r)&=&2\sqrt{r^2+1}\,\mathrm{arctan}\left(\sqrt{\dfrac{\r-r^2}{1+r^2}}\right)-\pi\sqrt{\r+1}+\pi\sqrt{r^2+1}-2\sqrt{r^2+1}\,\mathrm{arcsin}\left(\sqrt{\dfrac{\r-r^2}{\r+1}}\right)\nn
&=&\pi\sqrt{r^2+1}+\text{const.}\nn
\thus\dfrac{1}{\sqrt{h(r)}}&=&\dfrac{1}{r\pi}\dfrac{d}{dr}\pi\sqrt{r^2+1}\nn
\thus h(r)&=&1+r^2\quad\text{as expected}\nn
\eea

\newpage

\subsubsection{Numerical extraction using equation \eqref{exhr}}

We consider an oscillatory $h(r)$ to show that the extraction method is not limited to simple functions, and that the inversion reproduces every bump.

\begin{figure}[htbp]
\unitlength1cm \hfil
\begin{picture}(7,7)
 \epsfxsize=9cm \put(-1,0){\epsffile{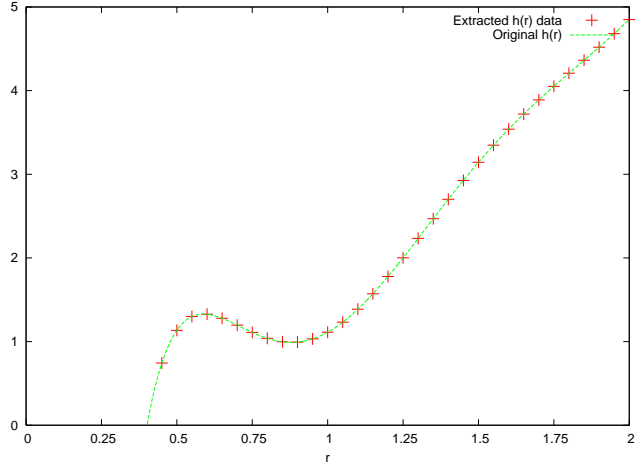}}
\end{picture}
\caption{Testing equation \eqref{exhr} for $h(r) = 1 + r^2 - \left[\dfrac{\mathrm{cos}(2.8r)}{r}\right]^2$}
\end{figure}

Notice that we can only extract down to $h(r)=0$ as this is the position of the event horizon.

\subsection{Extracting $f(r)$}

Now that we have a method of determining $h(r)$, we can use the original method described in section 2 using the path of null geodesics. The method will be exactly the same except for the modified equations of motion given in \eqref{nullstar}.

\paragraph{}By substituting in the modified $\dot{r}$, using the same substitutions as before, we get from Abel's equation \cite{Abel} that:

\be
q(V)=\dfrac{1}{\pi}\dfrac{\d}{\d V}\int^{V}_1\dfrac{\D\p\left(\displaystyle\sqrt{V_{max}}\right)\,\d V_{max}}{\displaystyle\sqrt{V-V_{max}}}\quad\text{where}\quad q(V)=-2\dfrac{\d r}{\d V}\dfrac{V}{\sqrt{f[V^{-1}(V)]\,h[V^{-1}(V)]}}
\ee

Sustituting the variables back into the equation as before, we finally get:
\begin{equation}
\label{star}
V^{\prime}(r) + \dfrac{1}{rI_1[V(r)]}\displaystyle\sqrt{\dfrac{V(r)}{h(r)}} = 0,\quad\text{where}\quad I_1[V] = -\dfrac{1}{\pi}\dfrac{\d}{\d V}\int^{\sqrt{V}}_1\dfrac{\alpha\Delta\phi(\alpha)\,\d\alpha}{\displaystyle\sqrt{V - {\alpha}^2}}
\end{equation}
As we know how to extract $h(r)$, we can solve equation \eqref{star} (e.g. by separation of variables) to find $V(r)$ and thus $f(r)$.

\subsubsection{Analytic extraction  using equation \eqref{star}}

Again, using the simplest case of pure AdS, we know that $\Delta\phi(\alpha) = \pi$ and $h(r) = 1+r^2$. We can now use \eqref{star} to solve for $V(r)$:
\begin{equation}
I_1[V] = -\dfrac{\d}{\d V}\int^{\sqrt{V}}_1\dfrac{\alpha\,\d\alpha}{\sqrt{V - {\alpha}^2}} = -\dfrac{1}{2\sqrt{V - 1}}
\end{equation}
\begin{equation}
\Rightarrow V^{\prime}(r) = -\dfrac{2\sqrt{V(r)\,(V(r) - 1)}}{r\sqrt{r^2 + 1}}
\end{equation}
Separation of variables gives \cite{integrals},
\begin{equation}
\mathrm{ln}\left[2\sqrt{V^2(r) - V(r)} + 2V(r) - 1\right] = \mathrm{ln}\left\lvert\dfrac{\sqrt{1+r^2} + 1}{\sqrt{1+r^2} - 1}\right\rvert
\end{equation}
\begin{equation}
\Rightarrow V(r)= \dfrac{1+r^2}{r^2} \Rightarrow f(r)=1+r^2\quad\text{as expected}
\end{equation}

\subsubsection{Numerical extraction using equation \eqref{star}}

Here we picked two different functions $h(r)$ and $f(r)$ to fully test the extraction.  

\begin{figure}[htbp]
\unitlength1cm \hfil
\begin{picture}(7,7)
 \epsfxsize=9cm \put(-1,0){\epsffile{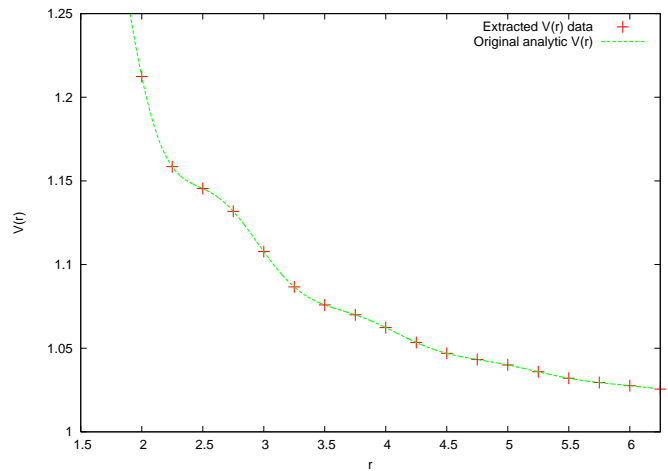}}
\end{picture}
\caption{Testing equation \eqref{star} for $V(r) = 1 + \dfrac{1}{r^2} - \left[\dfrac{\mathrm{cos}(2.8r)}{r^2}\right]^2$ and $h(r) = 1+r^2-\dfrac{1}{r^2}$}
\end{figure}

\paragraph{}As in section 2, we can only extract up to when $V(r)$ is a local maximum. There is no limiting in the extracting of $f(r)$ due to extracting $h(r)$ first as the radius of the event horizon is smaller than the radius of null circular orbits.

\paragraph{}In summary, we have found an analytic inversion method for extracting static spherically symmetric bulk spacetimes in three dimensions descibed by two functions, given information of the bulk-cone singularities and given the entanglement entropy of the two dimensional boundary CFT. Such spacetimes that can be extracted from this method include deformations of three dimensional AdS by a self-gravitating gas of radiation (so-called ``star'' geometries). In deriving the method, Abel's equation has been used twice on naively independent inversions. This might hint at a relationship between spacelike and null boundary-to-boundary geodesics in the context of this extraction. It was shown from graphical examples that any spacetime of this type can be extracted down to the local maximum of the potential for $f(r)$ and that the extraction of $h(r)$ does not change this limiting factor. 

\newpage
\section{Discussion}

In this paper we have given an analytic method for extracting bulk geometries given the location of singular correlators and the entanglement entropy in the boundary CFT in the context of the AdS/CFT correspondence. This was achieved in two particular types of geometry: Static spherically symmetric, asymptotically AdS spacetimes described by one function $f(r)$, and the same type of spacetime described by two independent functions $f(r)$ and $h(r)$. This was done by recognising linearity in the expression for $\D\p(\a)$ (and $\D t(\a)$), therefore allowing an analytic integral inversion.

\paragraph{}One limitatation in the extraction of $h(r)$ was the fact that the area of a minimal surface is only the proper length in a three dimensional spacetime. That means our analytic extraction given knowledge of the entanglement entropy is limited to the case of three dimensions (or a two dimensional boundary CFT). The natural progression would be to determine the area of higher dimensional minimal surfaces, and try to extract the metric from this, although the extraction method will be a lot more involved than the method laid out in this paper. Of course, the method for extracting $h(r)$ is not necessarily limited to three dimensions. Since the proper length is a physical quantity, intuitively there should be some physical quantity in the boundary field theory which is the dual of this quantity in the bulk. If that were found, $h(r)$ could be determined in any number of dimensions via the method shown. 

\paragraph{}In both cases where the metric is desribed by either one or two functions, the entire metric can be recovered if the spacetimes have a local maximum when $r\leqslant 0$. On the other hand, if the spacetimes admit null circular orbits, then the metric can only be determined down to the local maximum of $V(r)$. An extension to this work would be to look at cases where the spacetime geometry has less symmetry. This can by done by relaxing the spherical symmetry in one or more dimensions and by promoting the functions $f(r)$ and $h(r)$ to become time dependent. The ultimate goal would then be to completely extract the most general asymptotically AdS spacetime given boundary CFT data.

\section*{Acknowledgements}

I would like to thank Veronika Hubeny for her help in the construction of this paper and for her very useful insights into the beautiful world of general relativity and all of its rich texture.

\newpage

\startappendix

\Appendix{Null Geodesics}
\label{null}

For motion along geodesics, we can just consider the equatorial plane. So when calculating the equations of motion, we consider just one angular direction $\phi$.

\paragraph{}The equations of motion for $t$ and $\p$ yeild 2 constants of motion
\be
\dot{t}\,f(r) = E \quad\text{and}\quad r^{2}\dot{\p} = J,\quad\left(\dot{x} = \frac{\d x}{\d\lambda},\quad\text{where $\lambda$ is the affine parameter}\right) 
\ee
For null geodesics, $\d s = 0$, thus the equation of motion for r along null trajectories is given by
\be
\dot{r}^2 = E^2 - \dfrac{f(r)\,J^2}{r^2}
\ee
Dividing by $J$ overall, we can just consider the ratio $\a = \frac{E}{J}$ to define the null trajectories. Thus, we have the 3 equations of motion
\be
\dot{t} = \dfrac{\a}{f(r)},\quad\dot{\p} = \dfrac{1}{r^2},\quad\dot{r}^2 = {\a}^2 - \dfrac{f(r)}{r^2}
\ee

To calculate the angular separation, we use the fact that $\frac{\d\p}{\d r}=\frac{\dot{r}}{\dot{\p}}$ and thus

\be
\D\p(\a) = 2\int^{\infty}_{r_{min}(\a)}\frac{1}{r^2}\frac{1}{\sqrt{\a^2 - \frac{f(r)}{r^2}}}\,\d r
\ee
and similarly
\be
\D t(\a) = 2\int^{\infty}_{r_{min}(\a)}\frac{\a}{f(r)}\frac{1}{\sqrt{{\a}^2 - \frac{f(r)}{r^2}}}\,\d r
\ee

where the factor of 2 comes from the fact that the null trajectory goes from the boundary, down to a mimimun value of $r$, then back to the boundary at $r=\infty$

\Appendix{Derivation of equation \eqref{dadph}}

We can calculate the derivative of any function of the form $a(\a)=\int^{\infty}_{b(\a)}c(\a,r)\,dr$, where $b(\a)$ and $c(\a,r)$ are continuous, by using the definition of a derivative, i.e.
\be
\label{deriv}
\begin{split}
a^{\prime}(\alpha)=&\lim_{\delta\a\rightarrow\,0}\dfrac{\int^{\infty}_{b(\a+\delta\a)}c(\a+\delta\a,r)\,\d r-\int^{\infty}_{b(\a)}c(\a,r)\,\d r}{\delta\a}\\
=&-\,b^{\prime}(\a)\,c(\a,b(\a))+\int^{\infty}_{b(\a)}c^{\prime}(\a,r)\,\d r
\end{split}
\ee

This can now be applied to the definitions of $\D\p(\a)$ and $\D t(\a)$:
\be
\begin{split}
\a\,\dfrac{\d}{\d\a}\,\D t(\a)&=-\dfrac{2\a^2\,r_{min}^{\prime}(\a)}{r^2_{min}(\a)\,V(r_{min}(\a))\sqrt{\a^2-V(r_{min}(\a))}}+2\a\int^{\infty}_{r_{min}(\a)}\dfrac{\d}{\d\a}\left(\dfrac{\a}{r^2V(r)}\dfrac{1}{\sqrt{\a^2-V(r)}}\right)\,\d r\\
&=-\dfrac{2\,r_{min}^{\prime}(\a)}{r^2_{min}(\a)\sqrt{\a^2-V(r_{min}(\a))}}-2\int^{\infty}_{r_{min}(\a)}\dfrac{\a}{r^2(\a^2-V(r))^{\frac{3}{2}}}\,\d r\\
&=\dfrac{\d}{\d\a}\D\p(\a)\quad\text{using eq. \eqref{deriv}}\\ 
\end{split}
\ee

\Appendix{Spacelike Geodesics}
\label{spgeo}

For spacelike geodesics in with metric given by equation \eqref{frhr} we have:
\begin{equation}
-f(r)\,{\dot{t}}^2 + \dfrac{{\dot{r}}^2}{h(r)} + r^2{\dot{\phi}}^2 = \epsilon
\end{equation}
where $\epsilon > 0$ and will be normalised to 1 by redefining the affine parameter $\lambda$.
\paragraph{}To calcualte the proper length, we can consider the simplest case of static geodesics (i.e. $\dot{t} = 0$). Hence,
\begin{equation}
\begin{split}
&\quad\quad\dfrac{{\dot{r}}^2}{h(r)} + r^2{\dot{\phi}}^2 = 1\\
\Rightarrow\quad\dot{r} = &\sqrt{h(r)\left(1 - \dfrac{J^2}{r^2}\right)}\quad\text{by subtitution of $\dot{\phi}$}
\end{split}
\end{equation}
The definition of the proper length is given by:
\begin{equation}
\L(J) = \int{d}s = \int{d}\lambda = 2\int^{\rho_0}_{r_{min}(J)}\dfrac{dr}{\dot{r}},\quad\text{where}\quad\dot{r}\mid_{r_{min}(J)} = 0
\end{equation}

Where $\dot{r}\mid_{r_{min}(J)} = 0\thus r_{min}(J)=J$ or $h(r_{min}(J))=0$. But at $h(r)=0$ we have an event horizon, and static spacelike geodesics cannot extend inside an event horizon, so we use $r_{min}(J)=J$.  We have also introduced a maximum radius $r=\rho_0$ to regulate $\L$. This corresponds to the ultra-violet (UV) cut-off in the dual CFT and so is perfectly valid. Thus the integration region is $\rho_0\geq r\geq J$.

\paragraph{}Thus the final expression for the proper length is given by

\begin{equation}
\L(J) = 2\int^{\rho_0}_J\dfrac{2r\,\d r}{\sqrt{h(r)\,(r^2 - J^2)}}
\end{equation}

\end{document}